\documentclass[a4paper,12pt]{article}
\usepackage{eurosym}
\usepackage[dvips]{graphicx}
\usepackage{amsfonts}
\usepackage{mathrsfs}
\usepackage{graphicx} 
\usepackage{epsfig}
\usepackage{amsmath, amsthm}
\usepackage{amssymb}

\newcommand{\ba}{\begin{array}}
\newcommand{\ea}{\end{array}}

\newcommand{\beq}{\begin{equation}}
\newcommand{\eeq}{\end{equation}}
\newcommand{\ben}{\begin{enumerate}}
\newcommand{\een}{\end{enumerate}}
\newcommand{\bit}{\begin{itemize}}
\newcommand{\eit}{\end{itemize}}

\begin{document}
\title{Integrable Motion of Curves, Spin Equation and Camassa-Holm Equation}
\author{Assem Mussatayeva\footnote{Email: bekovaguldana@gmail.com},  \,           Tolkynay Myrzakul\footnote{Email: trmyrzakul@gmail.com}, \, Gulgassyl  Nugmanova\footnote{Email: nugmanovagn@gmail.com},  \,\\ Kuralay  Yesmakhanova\footnote{Email: kryesmakhanova@gmail.com} \, and Ratbay Myrzakulov\footnote{Email: rmyrzakulov@gmail.com}\\
\textsl{Eurasian International Center for Theoretical Physics and} \\ { Department of General \& Theoretical Physics}, \\ Eurasian National University,
Nur-Sultan, 010008, Kazakhstan
}
\date{}
\maketitle

\begin{abstract}
In the present paper, we investigate some geometrical properties of the Camass-Holm equation (CHE). We establish the geometrical equivalence between the CHE and the M-CIV  equation using a link  with the motion of  curves. We also show that these two equations are gauge equivalent each to other.
\end{abstract}

\section{Introduction}
In this paper, we will study the  geometry of  the Camassa-Holm equation (CHE) and its  integrable spin equivalent counterpart - the so-called M-CIV equation. The CHE is given by
\begin{equation}
u_{t}-u_{xxt}+3uu_{x}-2uu_{xx}-uu_{xxx}-\alpha u_{x}=0,  \label{1}
\end{equation}
where $u(x,t)$ is a real function, $\alpha\in R$. This equation  was firstly obtained in \cite{1} and later rediscovered by Camassa and Holm in \cite{2} as a model for shallow water waves. There
are many researchers investigated the CHE by different analytical
methods of solving the governing equation. In particular, it  was shown to be an
integrable equation admitting a bi-Hamiltonian formulation and a Lax pair \cite{3}. Due to its many remarkable properties, the CHE has attracted a lot of attention in the recent past.  In literature were  also constructed some  generalizations
or correlated equations of the CHE. One of  the generalisations is the following Dullin-Gottwald-Holm  equation
\begin{eqnarray}
q_{t} + uq_{x} + 2u_{x}q - \alpha u_{x} - \Gamma u_{xxx}&=&0, \label{2}\\
q-u+u_{xx}&=&0.  \label{3}
\end{eqnarray}
Note that the CHE can be rewrittten as
\begin{eqnarray}
q_{t}+2u_{x}q+uq_{x}&=&0,  \label{4}\\
q-u+u_{xx}&=&0,  \label{5}
\end{eqnarray}
where $u,q$ are some real functions.  Another interesting subclass of integrable systems is the so-called spin systems (equations). First example of integrable spin systems is the Heisenberg ferromagnet equation (HFE) of the form
\begin{equation}
{\bf A}_{t}+{\bf A}\wedge{\bf A}_{xx}=0,  \label{6}
\end{equation}
where 
\begin{equation}
{\bf A}=(A_{1},A_{2},A_{3}), \quad {\bf A}^{2}=1.  \label{7}
\end{equation}
Another  example of integrable spin systems is  the M-CIV equation. The M-CIV equation is given by
\begin{eqnarray}
{\bf A}_{xt}+(u{\bf A}_{x})_{x}+\frac{k}{4}{\bf A}\wedge{\bf A}_{x}+2({\bf A}_{x}\cdot{\bf A}_{t}+4u{\bf A}_{x}^{2}){\bf A}&=&0,  \label{8} \\
{\bf A}_{x}^{2}+4\beta^{2}(u-u_{xx})&=&0.  \label{9} 
\end{eqnarray}

In this paper, one of our goals is to study the relation between the CHE and the M-CIV equation and their connections with differential geometry of curves. 

The outline of the present paper is organized as follows. In Section 2, we
present  the M-CIV equation. Sections 3 and 4, the relations between the motion of space and plane curves, the CHE and the M-CIV equation is established. Then using this relation we found that the Lakshmanan (geometrical) equivalent counterpart of the M-CIV equation is the CHE. The gauge equivalence between the M-CIV equation and the CHE is mentioned  in Section 5. In Section 6, we present  the known peakon solutions of the CHE.  The dispersionless CHE is discussed in Section 7. The paper is concluded by comments and remarks in
Section 8.

\section{M-CIV equation}
Consider the M-CIV equation
\begin{eqnarray}
 [A,A_{xt}+(uA_{x})_{x}]-\frac{1}{\beta^{2}}A_{x}=0,\label{10}
\end{eqnarray}
where $\beta=const$ and 
\begin{eqnarray}
u=(1-\partial^{2}_{x})^{-1}\det{A_{x}}=\frac{1}{{\bf A}_{x}^{2}}\left[C-\int({\bf A}_{x}\cdot{\bf A}_{xt})dx\right], \quad {\bf A}=(A_{1}, A_{2}, A_{3}), \label{11}
\end{eqnarray}
\begin{eqnarray}
A=\left(\begin{array} {cc} A_{3}& A^{-} \\ 
A^{+}&-A_{3}\end{array}\right), \quad A^{\pm}=A_{1}\pm iA_{2}, \quad A^{2}=I, \quad {\bf A}^{2}=1.\label{12}
\end{eqnarray}
Its LR reads as
\begin{eqnarray}
\psi_{x}&=&U\psi,\label{13}\\
 \psi_{s}&=&V\psi \label{14},
\end{eqnarray}
where
\begin{eqnarray}
U&=&\left(\frac{\lambda}{4\beta}-\frac{1}{4}\right)[A,A_{x}],\label{15}\\
V&=&\left(\frac{1}{4\beta^{2}}-\frac{1}{4\lambda^{2}}\right)A+\left[\frac{1}{8\beta\lambda}-\frac{1}{8\beta^{2}}-\left(\frac{\lambda}{4\beta}-\frac{1}{4}\right)u\right][A,A_{x}]+\nonumber
\\
& + &\left(\frac{1}{2\lambda}-\frac{1}{2\beta}\right)[A,\frac{\beta}{2}A_{t}+\left(\frac{\beta u}{2}-\frac{1}{4\beta}\right)A_{x}]. \label{16}
\end{eqnarray}

\section{Motion of space curves intuced by the M-CIV equation}

Note that to the CHE corresponds the motion of plane curves. But in this section, we  study the "formally" more general case, namely, the motion of space curves induced by the  M-CIV
  equation.   
	\subsection{Variant-I}
	Consider a smooth space curve in $R^{3}$ given by 
\begin{eqnarray}
{\bf \gamma} (x,t): [0,X] \times [0, T] \rightarrow R^{3},\label{17}
\end{eqnarray} 
where $x$ is the arc length of the curve at each time $t$.  The corresponding  unit tangent vector ${\bf e}_{1}$,  principal normal vector ${\bf e}_{2}$ and binormal vector ${\bf e}_{3}$ are given by
\begin{eqnarray}
{\bf e}_{1}={\bf \gamma}_{x}, \quad {\bf e}_{2}=\frac{{\bf \gamma}_{xx}}{|{\bf \gamma}_{xx}|}, \quad {\bf e}_{3}={\bf e}_{1}\wedge {\bf e}_{2}, \label{48}
\end{eqnarray} 
respectivily. The   Frenet-Serret equation reads as 
 \begin{eqnarray}
\left ( \begin{array}{ccc}
{\bf  e}_{1} \\
{\bf  e}_{2} \\
{\bf  e}_{3}
\end{array} \right)_{x} = C
\left ( \begin{array}{ccc}
{\bf  e}_{1} \\
{\bf  e}_{2} \\
{\bf  e}_{3}
\end{array} \right)=
\left ( \begin{array}{ccc}
0   & \kappa_{1}     & \kappa_{2} \\
-\kappa_{1}& 0     & \tau  \\
-\kappa_{2}   & -\tau & 0
\end{array} \right)\left ( \begin{array}{ccc}
{\bf  e}_{1} \\
{\bf  e}_{2} \\
{\bf  e}_{3}
\end{array} \right), \label{49} 
\end{eqnarray}
where $\tau$,  $\kappa_{1}$ and $\kappa_{2}$ are   torsion,  geodesic curvature and  normal curvature of the curve, respectively. 
Integrable  deformation of the curves  are given by 
\begin{eqnarray}
\left ( \begin{array}{ccc}
{\bf  e}_{1} \\
{\bf  e}_{2} \\
{\bf  e}_{3}
\end{array} \right)_{x} = C
\left ( \begin{array}{ccc}
{\bf  e}_{1} \\
{\bf  e}_{2} \\
{\bf  e}_{3}
\end{array} \right),\quad
\left ( \begin{array}{ccc}
{\bf  e}_{1} \\
{\bf  e}_{2} \\
{\bf  e}_{3}
\end{array} \right)_{t} = G
\left ( \begin{array}{ccc}
{\bf  e}_{1} \\
{\bf  e}_{2} \\
{\bf  e}_{3}
\end{array} \right). \label{50} 
\end{eqnarray}
Here
\begin{eqnarray}
C &=&-\tau L_{1}+\kappa_{2}L_{2}-\kappa_{1}L_{3}=
\left ( \begin{array}{ccc}
0   & \kappa_{1}     & \kappa_{2}  \\
-\kappa_{1}  & 0     & \tau  \\
-\kappa_{2}    & -\tau & 0
\end{array} \right) ,\\
G &=&\omega_{1}L_{1}+\omega_{2}L_{2}-\omega_{3}L_{3}=
\left ( \begin{array}{ccc}
0       & \omega_{3}  & \omega_{2} \\
-\omega_{3} & 0      & \omega_{1} \\
-\omega_{2}  & -\omega_{1} & 0
\end{array} \right),\label{51} 
\end{eqnarray}
where
\begin{eqnarray}
 L_{1} = \begin{bmatrix}0&0&0\\0&0&-1\\0&1&0\end{bmatrix} , \quad
 L_{2} = \begin{bmatrix}0&0&1\\0&0&0\\-1&0&0\end{bmatrix} , \quad
 L_{3} = \begin{bmatrix}0&-1&0\\1&0&0\\0&0&0\end{bmatrix}
\end{eqnarray}
are basis elements of $so(3)$ algebra.
The compatibility condition of these equations is written as
\begin{eqnarray}
C_t - G_x + [C, G] = 0\label{52} 
\end{eqnarray}
or in elements   
 \begin{eqnarray}
\kappa_{1t}- \omega_{3x} -\kappa_{2}\omega_{1}+ \tau \omega_2&=&0, \label{25} \\ 
\kappa_{2t}- \omega_{2x} +\kappa_{1}\omega_{1}- \tau \omega_3&=&0, \label{26} \\
\tau_{t}  -    \omega_{1x} - \kappa_{1}\omega_2+\kappa_{2}\omega_{3}&=&0.  \label{27} \end{eqnarray}
Our next step is the following identification
 \begin{eqnarray}
{\bf A}\equiv {\bf e}_{1}. \label{28} 
\end{eqnarray}
We now make  the following "formal" assumption    
\begin{eqnarray}
\kappa_{1}=i, \quad \kappa_{2}=\lambda(q-1), \quad \tau=-i\lambda(1+q), \label{57} 
\end{eqnarray}
where $q=-0.5(\kappa_{2}-i\tau)$ and $r=0.5(\kappa_{2}+i\tau)$ are some  functions,  $\lambda=const$.  Then we have 
\begin{eqnarray}
\omega_{1} & = &i[(0.5\lambda^{-1}-\lambda u)(q+1)+0.5\lambda^{-2}(u_{x}+u_{xx})],\label{58}\\ 
\omega_{2}&=& [(0.5\lambda^{-1}-\lambda u)(q+1)+0.5\lambda^{-1}(u_{x}+u_{xx})], \label{59} \\
\omega_{3} & = &i[0.5\lambda^{-2}-u-u_{x}].      \label{60}
\end{eqnarray}
Eqs.(\ref{25})-(\ref{27}) give us the following equations for $q, u$: 
\begin{eqnarray}
q_{t}+2u_{x}q+uq_{x}&=&0, \label{33} \\
q-u+u_{xx}&=&0, \label{62} \label{34}
\end{eqnarray}
which is the CHE. As it is well-known that the CHE is integrable. Its LR has the form
\begin{eqnarray}
\psi_{xx}&=&(0.25+\lambda^{2}q)\psi, \label{35} \\
\psi_{t}&=&(0.5\lambda^{-2}-u)\psi_{x}+0.5u_{x}\psi. \label{36}
\end{eqnarray}

So, we have  proved  the  Lakshmanan (geometrical) equivalence between the M-CIV  equation  (\ref{10}) and the CHE (\ref{33})-(\ref{34}). 

\subsection{Variant-II}

In this section, let us we assume that $\kappa_{2}=0$.  Then matrices $C$ and $D$ take the form
\begin{eqnarray}
C &=&-\tau L_{1}-\kappa_{1}L_{3}=
\left ( \begin{array}{ccc}
0   & \kappa_{1}     & 0  \\
-\kappa_{1}  & 0     & \tau  \\
0    & -\tau & 0
\end{array} \right), \label{37}\\
G &=&\omega_{1}L_{1}+\omega_{2}L_{2}-\omega_{3}L_{3}=
\left ( \begin{array}{ccc}
0       & \omega_{3}  & \omega_{2} \\
-\omega_{3} & 0      & \omega_{1} \\
-\omega_{2}  & -\omega_{1} & 0
\end{array} \right).\label{38} 
\end{eqnarray}
At the same time, the equations (\ref{25})-(\ref{27}) read as
\begin{eqnarray}
\kappa_{1t}- \omega_{3x} + \tau \omega_2&=&0, \label{39} \\ 
\omega_{2x} -\kappa_{1}\omega_{1}+ \tau \omega_3&=&0, \label{40} \\
\tau_{t}  -    \omega_{1x} - \kappa_{1}\omega_2&=&0.  \label{41} \end{eqnarray}
We now again take the identification (\ref{28}). The M-CIV equation  can rewrite as
\begin{eqnarray}
A_{xt}+(uA_{x})_{x}+\frac{k}{4}[A,A_{x}]+0.5(A_{t}A_{x}+A_{x}A_{t}+2uA_{x}^{2})A=0.  \label{42} \end{eqnarray}
In the vector form it takes the form
\begin{eqnarray}
{\bf A}_{xt}+(u{\bf A}_{x})_{x}+\frac{k}{4}{\bf A}\wedge{\bf A}_{x}+2({\bf A}_{x}\cdot{\bf A}_{t}+4u{\bf A}_{x}^{2}){\bf A}=0.  \label{43} \end{eqnarray}
As result, we have the following equations
\begin{eqnarray}
\omega_{2x}+\tau\omega_{3}+u\kappa_{1}\tau+\beta\kappa_{1}&=&0, \label{44} \\ 
\omega_{3x}-\tau\omega_{2}+(u\kappa_{1})_{x}&=&0, \label{45} \\ 
\omega_{1}-\frac{\omega_{2x}+\tau\omega_{3}}{\kappa_{1}}&=&0,   \label{46} \end{eqnarray}
so that we get
\begin{eqnarray}
\kappa_{1t}+(u\kappa_{1})_{x}=0. \label{47}
\end{eqnarray}
In this case we have
\begin{eqnarray}
tr(A_{x}^{2})=2{\bf A}_{x}^{2}=-8\beta^{2} q=2\kappa_{1}^{2}  \label{48} \end{eqnarray}
or
\begin{eqnarray}
q=-\frac{1}{4\beta^{2}}\kappa_{1}^{2}.  \label{49} \end{eqnarray}
Substituting all these formulas into the equation (\ref{47}) gives us the CHE (\ref{33})-(\ref{34})
\begin{eqnarray}
q_{t}+2u_{x}q+uq_{x}=0. \label{50}
\end{eqnarray}

\section{Motion of plane curves intuced by the M-CIV equation}
As we mentioned above, to the CHE corresponds an integrable motion of plane curves. For this reason, in this section, let us we assume that $\kappa_{2}=\tau=\omega_{1}=\omega_{2}=0$ that is we consider the plane curves.   Then matrices $C$ and $D$ take the form
\begin{eqnarray}
C =\left ( \begin{array}{cc}
0   & \kappa_{1}   \\
-\kappa_{1}  & 0  
\end{array} \right) ,\quad 
G =
\left ( \begin{array}{cc}
0       & \omega_{3}  \\
-\omega_{3} & 0     
\end{array} \right).\label{51} 
\end{eqnarray}
In this case, Eqs. (\ref{25})-(\ref{27}) take the form
\begin{eqnarray}
\kappa_{1t}- \omega_{3x} =0.  \label{52} 
\end{eqnarray}
We now again take the identification (\ref{28}). Then we have
\begin{eqnarray}
\omega_{3}=u\kappa_{1}.  \label{53} 
\end{eqnarray}
We now  assume that
\begin{eqnarray}
q=u-u_{xx}=-\frac{1}{4\beta^{2}}\kappa_{1}^{2}.  \label{54} \end{eqnarray}
Then the equation (\ref{52}) gives 
\begin{eqnarray}
q_{t}+2u_{x}q+uq_{x}=0
\end{eqnarray}
that is again the CHE.

\section{Gauge equivalence}
In the previous section, we have established the geometrical equivalence between the M-CIV equation and the CHE. Here we note that these equations is gauge equivalent each to other \cite{R99}.

\section{Peakon solutions}
It is well-known that the CHE has the so-called  peaked soliton solutions or  peakons for short. The $N$-peakon solution of the CHE has the  form
\begin{eqnarray}
u(x, t) = \sum_{j=1}^{N}m_{j}(t)e^{-|x-x_{j}(t)|},
\end{eqnarray}
where $x_{j}(t)$ and $m_{j}(t)$ are the positions and the amplitudes. These functions of the well-known peakon solutions  satisfy the following set of equations
\begin{eqnarray}
\dot{x}_{k}& =& \frac{\partial H}{\partial m_{k}}=\sum_{j=1}^{N}m_{j}e^{-|x_{k}-x_{j}|},\\
\dot{m}_{k}& =& -\frac{\partial H}{\partial x_{k}}=\sum_{j=1}^{N}m_{k}m_{j}sgn(x_{k}-x_{j})e^{-|x_{k}-x_{j}|},
\end{eqnarray}
where $H$ is the Hamiltonian of the form
\begin{eqnarray}
H=\frac{1}{2}\sum_{k, j=1}^{N}m_{k}m_{j}e^{-|x_{k}-x_{j}|}.
\end{eqnarray}
Similarly, we can construct the peakon solutions of the M-CIV equation which leave for the our next papers.

\section{Dispersionless CHE}
One of interesting subclass of integrable systems is the so-called dispersionless equations. It is well-known that  the CHE has  the dispersionless counterpart. Consider the CHE in the form
\begin{equation}
\epsilon u_{t}-\epsilon^{3}u_{xxt}+3\epsilon uu_{x}-2\epsilon^{2}uu_{xx}-\epsilon^{3}uu_{xxx}=\alpha \epsilon u_{x}.  \label{1}
\end{equation}
This equation we can rewrite as
\begin{equation}
 u_{t}-\epsilon^{2}u_{xxt}+3 uu_{x}-2\epsilon uu_{xx}-\epsilon^{2}uu_{xxx}=\alpha  u_{x}.  \label{1}
\end{equation}
Hence we obtain the following equation
\begin{equation}
 u_{t}+3 uu_{x}-\alpha  u_{x}=0.  \label{1}
\end{equation}
This is the dispersionless CHE \cite{1205.5831}.
\section{Conclusions}
 
In this paper, we have established the relation between the M-CIV  equation (10) and the CHE (33)-(34). We have shown that the M-CIV equation (10) and the CHE (33)-(34) is the geometrically equivalent each to other. Also the gauge  equivalence between these equations is shown. 
Our results are significant for the deep understand
of integrable spin systems and their relations with differential geometry of curves and surfaces (see e.g. \cite{R13}-\cite{R99}) in the peakon sector of integrable systems. 
 
 \section{Acknowledgements}
This work was supported  by  the Ministry of Edication  and Science of Kazakhstan under
grants 0118ะส00935 and 0118ะส00693.

 \end{document}